\documentclass[twocolumn,showpacs,preprintnumbers,amsmath,amssymb,showkeys]{revtex4}
\usepackage{graphicx}
\usepackage{subfigure}
\usepackage{epsfig}
\usepackage{dcolumn}
\usepackage{bm}
\usepackage{amsmath,amssymb}

\begin{document}

\title{Fluctuation theorems in inhomogeneous media under coarse graining}

\author{Sourabh Lahiri}
\email{sourabhlahiri@gmail.com}
\altaffiliation[Presently at ]{School of Physics, Korea Institute for Advanced Study.}
\author{Shubhashis Rana}
\author{A. M. Jayannavar}
\affiliation{Institute of Physics, Sachivalaya Marg, Bhubaneswar 751005, India}

\begin{abstract}
We compare the fluctuation relations for  work and entropy  in  underdamped and overdamped
 systems, when the friction coefficient of the medium is space-dependent. We find that these relations
remain unaffected in both cases. We have restricted ourselves to Stratonovich discretization scheme for the overdamped case.
\end{abstract}

\maketitle

\newcommand{\nwc}{\newcommand}
\nwc{\vs}{\vspace}
\nwc{\hs}{\hspace}
\nwc{\la}{\langle}
\nwc{\ra}{\rangle}
\nwc{\lw}{\linewidth}
\nwc{\nn}{\nonumber}

\nwc{\pd}[2]{\frac{\partial #1}{\partial #2}}
\nwc{\zprl}[3]{Phys. Rev. Lett. ~{\bf #1},~#2~(#3)}
\nwc{\zpre}[3]{Phys. Rev. E ~{\bf #1},~#2~(#3)}
\nwc{\zpra}[3]{Phys. Rev. A ~{\bf #1},~#2~(#3)}
\nwc{\zjsm}[3]{J. Stat. Mech. ~{\bf #1},~#2~(#3)}
\nwc{\zepjb}[3]{Eur. Phys. J. B ~{\bf #1},~#2~(#3)}
\nwc{\zrmp}[3]{Rev. Mod. Phys. ~{\bf #1},~#2~(#3)}
\nwc{\zepl}[3]{Europhys. Lett. ~{\bf #1},~#2~(#3)}
\nwc{\zjsp}[3]{J. Stat. Phys. ~{\bf #1},~#2~(#3)}
\nwc{\zptps}[3]{Prog. Theor. Phys. Suppl. ~{\bf #1},~#2~(#3)}
\nwc{\zpt}[3]{Physics Today ~{\bf #1},~#2~(#3)}
\nwc{\zap}[3]{Adv. Phys. ~{\bf #1},~#2~(#3)}
\nwc{\zjpcm}[3]{J. Phys. Condens. Matter ~{\bf #1},~#2~(#3)}
\nwc{\zjpa}[3]{J. Phys. A: Math theor  ~{\bf #1},~#2~(#3)}



\section{Introduction}

The last couple of decades have observed a steadily growing interest in the field of systems at mesoscopic scales,
 thanks to the growing understanding of machines and engines with smaller dimensions. This 
has led to the area of stochastic thermodynamics which provides a framework for extending notions of classical 
thermodynamics to small systems wherein concepts of work, heat, and entropy are extended to the level of 
individual trajectories during nonequilibrium processes (ensembles). Research in this area  has given
 birth to a group of exact and powerful theorems that dictate the behavior of such systems.
 They are commonly referred to as the \emph{fluctuation theorems} (FTs)
\cite{eva02,har07,rit03,rit06,kur07,eva93,eva94,gal95,jar97,zon02,zon04,dha04,sei05,sei08,cro98,cro99,sei12},
 and these theorems are valid even far from equilibrium, a feat that is beyond the scope of the well-established 
linear response theory. The theorems provide stringent restrictions on the probabilities of phase space 
trajectories in which second law is transiently ``violated''. They show that at the level where fluctuations
 are comparable to the relevant energy exchanges of the system, one needs to replace the associated quantities
 in the statement of the second law by their 
\emph{averages}: $\la W\ra\ge\Delta F$ or $\la \Delta s_{tot}\ra\ge 0$ \cite{jar97,sei05,sei08}. Here the angular
 brackets represent the ensemble average. Thus, they in essence uphold the
  second law, even at the mesoscopic level, however, for the average properties.

The Crooks Fluctuation theorem (CFT) for heat states that the ratio of the probabilities of forward trajectory
 and the corresponding reverse trajectory for given initial states is given by \cite{cro98,cro99}
\begin{equation}
 \frac{P[X|x_0]}{\tilde{P}[\tilde{X}|x_{\tau}]}=e^{\beta Q}.
\label{mic_rev}
\end{equation}
Here, X is the short form of the phase space trajectory along the forward process $x_0,x_1,...,x_{\tau}$
generated by the protocol $\lambda(t)$. $x_i$ represents the phase space point at time $t_i$. $\tilde{X}$
is the corresponding reverse trajectory generated by the time reversed protocol $\lambda(\tau-t)$, where $\tau$ 
is the time of observation. $x_0$ is a given initial state of the forward process. The reverse process 
begins from the state $\tilde{x}_{\tau}$, which is the time-reversal of the final state $x_{\tau}$ of 
the forward process. 

Using CFT, several other theorems like the Jarzynski equality and entropy production FT,
 can be easily derived \cite{cro98,cro99}.

In this paper, we study the validity of these FTs in the presence of coarse-graining,  when we transform the underdamped Langevin equation to the overdamped one, in the limit of high friction.
We find that a prominent difference in the analysis is observed between the overdamped (coarse-grained) and the
 underdamped systems, when the friction coefficient is space-dependent\cite{san82,jay95,jay01,lub07}. 
It should be noted that space-dependent friction does not alter the equilibrium state. However, Langevin dynamics of the system gets modified especially for the
overdamped case. There are several physical systems wherein friction is space-dependent 
(see \cite{lub07} and the references therein).

\section{CROOKS THEOREM IN PRESENCE OF SPACE-DEPENDENT FRICTION}
In the presence of space-dependent friction $\gamma(x)$, the equation of motion of the underdamped
system of mass m moving in a time-dependent potential $U(x,t)$ is given by
\begin{equation}
 m\dot v=-\gamma (x)v -U'(x,t)+\sqrt{2\gamma (x) T}\xi(t).
\label{L-ud}
\end{equation}
 Note that the above equation contains multiplicative noise term.
Here, T is the temperature of the bath, while $\xi(t)$ is the delta-correlated Gaussian noise with zero mean:
$\la \xi(t)\ra=0; \la\xi(t)\xi(t')\ra=\delta(t-t')$. The overhead dot denotes time-derivative, whereas prime 
represents space derivative. Eq. (\ref{L-ud}) has been derived microscopically by invoking system and 
bath coupling \cite{jay95,jay01}. It is shown that the high damping limit of eq. (\ref{L-ud}) is not equivalent to 
ignoring only inertial term \cite{san82,jay95,jay01,lub07}. The detailed treatment leads to  an extra term  that is crucial for system
 to reach equilibrium state in absence of time-dependent perturbations (see eq. \ref{eq:over} below). 

 Roughly speaking, this happens in the overdamped case because the random forces $\xi(t)$ appear as delta-function pulses that cause jumps in $x$. It then becomes unclear what value of $x$ must be provided in the argument of the function $g$, because the value of the position at the time the delta-peak appears becomes undefined \cite{van}. It does not converge to a unique value even in the limit of small time step $\Delta t$. In fact, we can plug in any value of position in-between $x(t)$ (position before the jump) and $x(t+\Delta t)$ (position after the jump). These different values of position lead to different discretization schemes. The case is simpler in case of underdamped Langevin equation. There, the jumps are caused in the velocities, while the position is a much smoother variable (being an integral over the velocities). In other words, it does not feel the noise as delta peaks, but instead as a more well-behaved function. In that case, in the limit of small $\Delta t$, the argument of $g$ is given by the unambiguous value $x(t)$. Thus, in this case, an update in the values of $x$ and $v$ will be unique in each time step.

Let us now check the validity of CFT
in both the underdamped and overdamped cases. 

\subsection{Underdamped case}
At first we want to calculate the ratio of path probabilities between forward and reverse process. 
In a given process, let the evolution of the system in phase space be denoted by the phase space
 trajectory $X(t)\equiv \{x_0,x_1,\cdots,x_\tau\}$. Here, $x_k$ represents the phase point at 
time $t=t_k$. In general, the phase point includes both the position and the velocity coordinates
 of the system. In the overdamped case, however, it would consist of the position coordinate only.
 Now, a given path $X(t)$, for a given initial point $x_0$, would be fully determined if the  sequence
 of noise terms for the entire time of observation is available ( this happens because there is no unambiguity in either the positions or the velocities, while updating their values by using the underdamped Langevin equation, as discussed above):
 $\bm{\xi} \equiv \{\xi_0,\xi_1,\cdots,\xi_{\tau-1}\}$. 
The probability distribution of $\xi_k$ is given by
\begin{equation}
P(\xi_k) \propto e^{-\xi_k^2dt/2}.
\end{equation}
Therefore, the probability of obtaining the sequence $\bm{\xi}$ will be \cite{dha04,pel06}
\begin{equation}
P[\bm{\xi(t)}] \propto \exp\left[-\frac{1}{2} \int_0^\tau\xi^2(t) dt\right].
\label{eq:P[eta]}
\end{equation}
Now, from the probability $P[\bm{\xi(t)}]$ of the path $\bm{\xi(t)}$ in noise space, we can obtain
 the probability $P[X(t)|x_0]$. These two probability functionals are related by the Jacobian 
$|\frac{\partial \xi}{\partial x}|$. Thus, we can as well write \cite{dha04}
\begin{equation}
P[X(t)|x_0] \propto \exp\left[-\frac{1}{2} \int_0^\tau\xi^2(t) dt\right],
\label{eq:P[X]}
\end{equation}
where the proportionality constant is different from that in eq. (\ref{eq:P[eta]}). In
 eq. (\ref{eq:P[X]}), we then substitute the expression for $\xi(t)$ from the Langevin equation (Eq.(\ref{L-ud})):
\begin{equation}
P[X(t)|x_0] \propto \exp\left[-\frac{1}{4} \int_0^\tau dt\frac{(m\dot v + U'(x,t) + \gamma(x) v)^2}{\gamma(x)T}\right].
\end{equation}
For the reverse process, $v\to -v$, but the Jacobian is same. The ratio of  probability 
of the forward to the reverse path can be readily shown to be\cite{dha04,ast06}
\begin{small}
 \begin{eqnarray}
&&  \frac{P[X(t)|x_0]}{\tilde P[\tilde X(t)|\tilde x_\tau]}\nn\\
&& = \frac{\exp\left[-\int_0^\tau dt(m\dot v + U'(x,t) + \gamma(x) v)^2/4\gamma(x)T\right]}
        {\exp\left[-\int_0^\tau dt(m\dot v + U'(x,t) - \gamma(x) v)^2/4\gamma(x)T\right]}\nn\\
&& =\exp\left[-\int_0^\tau dt\frac{4m\gamma(x)\dot v v+4U'(x,t)\gamma(x)v}{4\gamma(x)T}\right]\nn\\
&& =\exp\left[-\beta \int_0^\tau dt\left(m\dot v v+U'(x,t)v\right)\right]\nn\\
&& =e^{\beta Q},
\label{CrooksFT}
\end{eqnarray}
\end{small}
where Q is the heat dissipated by the system into the bath, defined as

\begin{eqnarray}
Q && \equiv \int_0^\tau \{\gamma(x)v-\sqrt{2\gamma(x)T}\xi(t)\}v~dt \nn\\
&&= -\int_0^\tau \left\{m\dot{v}+U'(x,t)\right\}v~dt.
\end{eqnarray}
This definition follow from the stochastic energetics developed by Sekimoto \cite{sek97,sek}
 from the definition of First Law using Langevin dynamics . Eq.(\ref{CrooksFT}) is the celebrated CFT, 
from which several FT follow.

\subsection{ Integral and detailed fluctuation theorems}

We have,
\begin{equation}
\frac{P[X(t)|x_0]}{\tilde P[\tilde X(t)|\tilde x_\tau]} = e^{\beta Q},
\end{equation}
where $Q$ is the heat dissipated, as obtained from the first law. 
Multiplying by the ratio of the initial equilibrium distributions, for forward and reverse processes,
 namely by $p_0(x_0)/p_1(x_\tau)$, we get \cite{cro98}
\begin{align}
  \frac{P[X(t)|x_0]p_0(x_0)}{\tilde P[\tilde X(t)|\tilde x_\tau]p_1(x_\tau)} &= \frac{P[X]}{\tilde P[\tilde X]} =e^{\beta Q} \cdot \frac{e^{-\beta E_0}}{Z(\lambda_0)}\cdot \frac{Z(\lambda_\tau)}{e^{-\beta E_\tau}} \nonumber\\
  &=  e^{\beta(Q+\Delta E-\Delta F)} = e^{\beta(W-\Delta F)}.
\label{JE_traj1}
\end{align}

We have used the expression for equilibrium initial distribution $p_0(x_0)=\frac{e^{-\beta E_0}}{Z(\lambda_0)}$
and $p_1(x_{\tau})=\frac{e^{-\beta E_{\tau}}}{Z(\lambda_{\tau})}$. Here, $\Delta E\equiv E_\tau-E_0$, and we have made use of the relation $Z=e^{-\beta F}$, between 
the partition function and the free energy. $Z(\lambda_0)$ and $Z(\lambda_\tau)$ are the partition
 functions corresponding to the protocol values at the initial time and the final time, respectively. 
In the final step, the first law for the work done on the system, $W=Q+\Delta E$, has been invoked.
 The above relation can be readily converted to the Crooks work theorem \cite{cro99}, given by 
\begin{equation}
\frac{P(W)}{\tilde P(-W)} =e^{\beta(W-\Delta F)}.
\end{equation}
Here, $P(W)$ is the probability of  work done $W$ on the system in the forward process. $\tilde P(-W)$
is the probability of $W$ amount of work extracted from the system in the reverse process.
By cross-multiplication and integration over $W$, we get the Jarzynski equality \cite{jar97}:
\begin{equation}
\left<  e^{-\beta W}\right> = e^{-\beta\Delta F}.
\label{JE}
\end{equation}

If the initial distributions for the forward and reverse processes are not equilibrium ones 
and $p_1(x_{\tau})$ is the solution of the Fokker-Planck Equation at the final time $\tau$
of the forward process, we get, instead of eq. (\ref{JE_traj1}), the relation \cite{sei05,sei08}
\begin{equation}
\frac{P[X]}{\tilde P[\tilde X]} =  e^{\beta Q+\ln (p_0(x_0)/p_1(x_\tau))} = e^{\Delta s_{tot}}.
\label{asd}
\end{equation}

We then arrive at the relations for change of total entropy $\Delta s_{tot}$ which is nothing but sum of 
change of system entropy $\Delta s_{sys}=\ln (p_0(x_0)/p_1(x_\tau))$ (in the units of Boltzmann constant 
$k_B$) and entropy production in the bath
$s_B=\beta Q$.
\begin{equation}
 \Delta s_{tot}=\ln (p_0(x_0)/p_1(x_\tau))+ \beta Q.
\end{equation}
 From Eq.(\ref{asd}) integral fluctuation theorem follows, which hold for all times, namely, 
\begin{align}
\left<  e^{-\Delta s_{tot}}\right> = 1.
\label{IFT}
  \end{align}

From the integral forms of the fluctuation theorems, given by eqs. (\ref{JE}) and (\ref{IFT}), 
using Jensen's inequality we easily obtain the second law inequalities \cite{jar97,sei08}
\begin{align}
\langle W\rangle &\ge \Delta F; \\
\left<\Delta s_{tot}\right> &\ge 0.
\end{align} 
Thus, in the underdamped limit second law retains same form for a system in presence of
 space-dependent friction. This completes our treatment for some FTs in the underdamped 
case for a particle moving in space dependent friction.
\begin{figure}
\vspace{0.5cm}
\centering
\includegraphics[width=\lw]{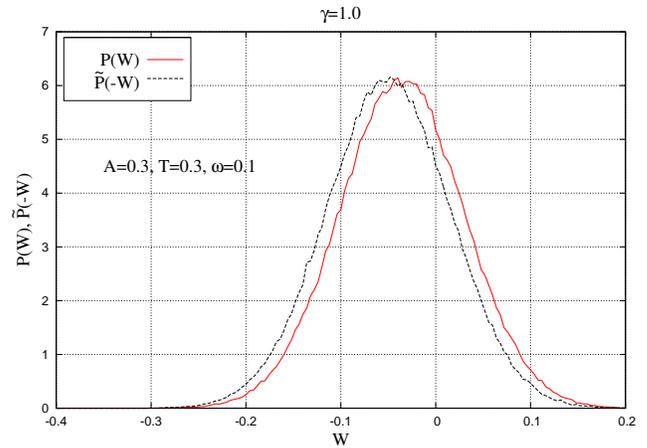}
\caption{ Transient work distribution for underdamped case }
\label{u-w0}
\end{figure}

\begin{figure}
\vspace{0.5cm}
\centering
\includegraphics[width=\lw]{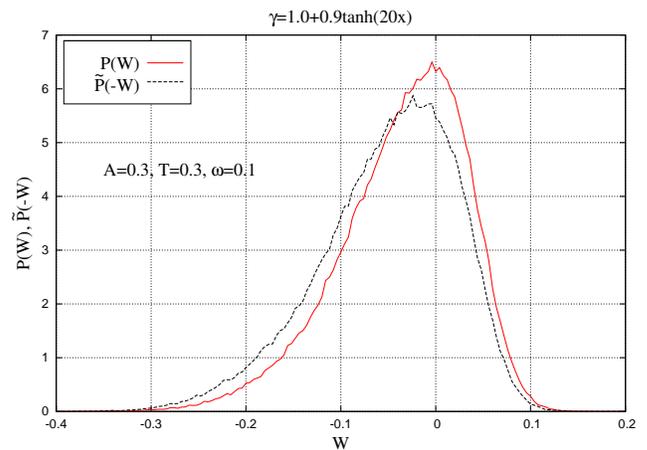}
\caption{Transient work distribution for underdamped case with space dependent friction }
\label{u-wx}
\end{figure}

Above exact FTs do not give any information  about probability distribution of work $(P(W))$, entropy $P(\Delta s_{tot})$
etc. These distributions depend crucially on the specific problem being investigated.

Here, we study these distributions for the case of driven particle in harmonic trap. Apart from verifying FTs we
also see how the space dependent friction modifies the distribution of $(P(W))$, and  $P(\Delta s_{tot})$
as compared to the particle moving in a space independent frictional coefficient $\gamma$ (which is the
 space average of $\gamma(x)$). The underdamped Langevin equation is given by
\begin{equation}
  m\dot v=-\gamma (x)v -k x +A \sin(\omega t)+\sqrt{2\gamma (x) T}\xi(t).
\end{equation}
$A \sin(\omega t)$ is driven sinusoidal force of frequency $\omega$ and amplitude $A$. For this model 
analytical solution can be obtained for space independent case only for both overdamped and underdamped 
case \cite{cil07,mam10}.

For simplicity in our study, we restrict ourselves to two cases of space dependent friction 
(i) $\gamma(x)=\gamma=$ constant (ii) $\gamma(x)=\gamma+c \tanh(\alpha x)$

In fig.(\ref{u-w0}) we have plotted the transient work distribution obtained after driving a system 
for one-fourth of a cycle for forward $(P(W))$ and corresponding reverse $(\tilde P(-W))$ protocol.
Initially the system is equilibrated at appropriate initial values of protocol for forward and reverse process.
In all our simulations, we have used the Heun's method of numerical integration \cite{man},
 and have generated $\sim 10^5$ realizations. Implementing the Heun's method tantamounts
 to using the Stratonovich discretization scheme \cite{sta}. Henceforth we have used all the quantities 
in dimensionless form and taken k=1, m=1 and $\gamma=1$. For case (i), both distributions are Gaussian nature, and they cross each other at $\Delta F=-0.044$, which is the free energy difference over 
 one-fourth cycle. This is obtained numerically from $\left<  e^{-\beta W}\right> = e^{-\beta\Delta F}$,
while theoretically we have $\Delta F=-0.045$. This is well within our numerical accuracy.  

In fig(\ref{u-wx}) we have plotted the same for space dependent friction $\gamma(x)=1.0+0.9 \tanh(20 x)$.
Here the distributions are non-Gaussian but the crossing point is same as in space independent case. This is because 
the equilibrium distribution remain same in both cases.

\begin{figure}
\vspace{0.5cm}
\centering
\includegraphics[width=\lw]{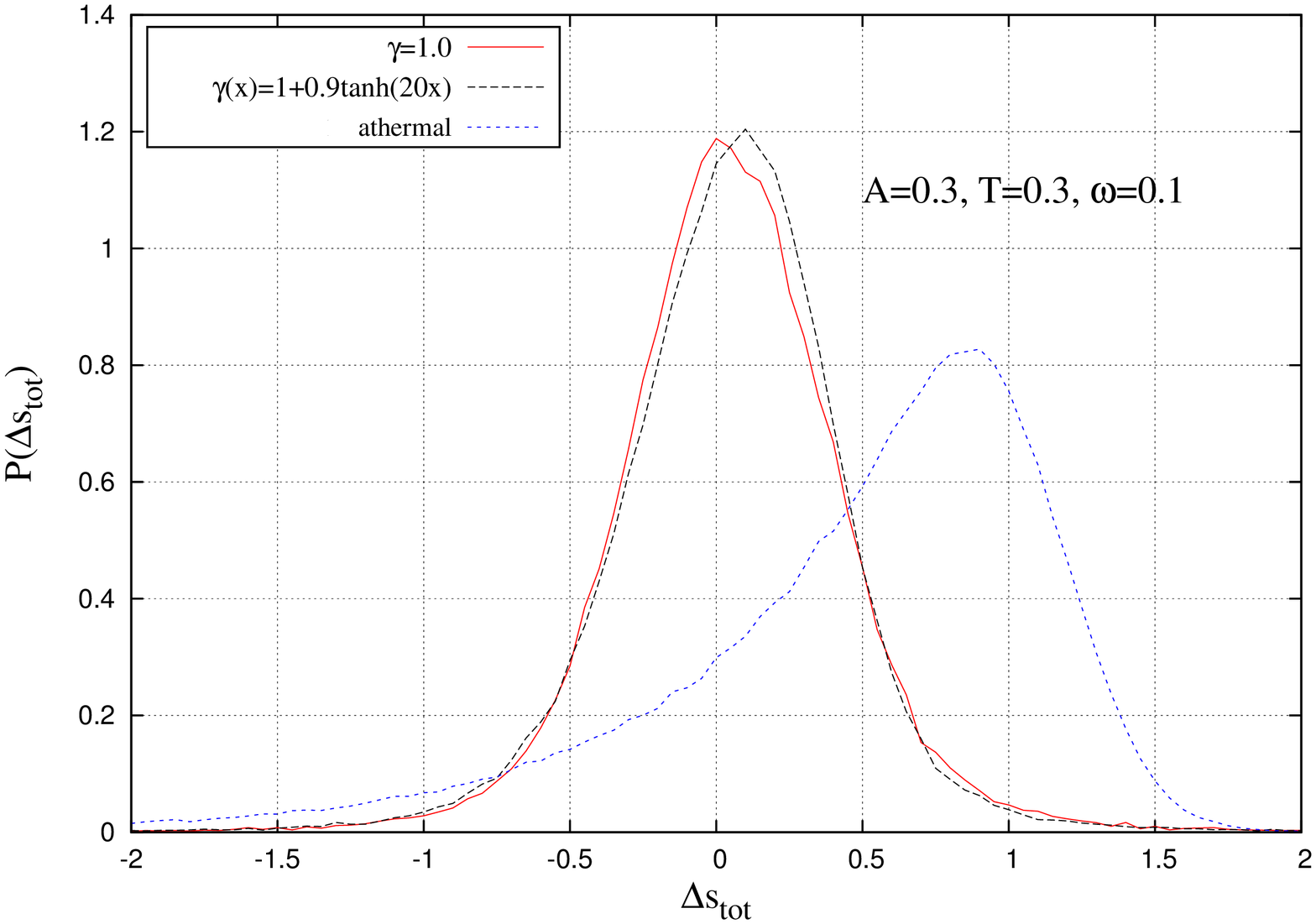}
\caption{Distribution for total entropy production for underdamped dynamics}
\label{u-s}
\end{figure}
In fig.(\ref{u-s}) we have plotted distribution of total entropy production of a driven Brownian particle
confined in a harmonic trap for one-fourth cycle. Here the underlying dynamics is underdamped and
 we find that the distribution is Gaussian for space independent case, while it is non-Gaussian for
space dependent case. If we take particle to be initially equilibrated at different temperature $T=0.1$
and then connected instantaneously to the given bath of temperature $T=0.3$, and driven by same external
 force (i.e, for athermal case), we find that the distribution of total entropy production is non-Gaussian
even for space independent case. This is consistent with the results in \cite{sah09}. Numerically we find
 $\la e^{-\Delta s_{tot}}\ra=1.002$ which is well within our numerical accuracy. In all these distributions,
 we find that there is a finite weight for realizations having $W<\Delta F$ and $\Delta s_{tot}<0$, although the 
mean values follow the second law inequalities. These realizations are called transient Second Law
violating trajectories. This finite weight is necessary to satisfy the fluctuation theorems \cite{mam11}.

After establishing the well known FTs in the underdamped case, we turn our attention to the overdamped
dynamics of the particle, in a space-dependent frictional medium.  Going to the 
overdamped regime implies coarse-graining. Instead of evolution in full phase space (coordinates and momenta),
we restrict the evolution of the system to the position space only. This is equivalent to 
ignoring the information contained in the velocity variables. 

\subsection{ Overdamped case}
The treatment of overdamped case is more subtle and a proper methodology must be followed.  In order to obtain a unique Fokker-Planck equation (which is needed for a unique equilibrium distribution), the overdamped Langevin equation must be modified, depending on the discretization process that is being used. It can be written as
\begin{eqnarray}
\dot x &&=f(x,t)+g(x)\xi(t)\nn\\
&&=-\Gamma(x)U'(x,t)+(1-\alpha)g(x)g'(x) + g(x)\xi(t).\nn\\
\label{eq:over}
\end{eqnarray}
For detail we refer to \cite{lub07}.  Such ambiguity of discretization process does not arises 
in the underdamped case as discussed in detail in \cite{van,man12}. 
Here, $g(x)=\sqrt{2T\Gamma(x)}=\sqrt{2T/\gamma(x)}$. $\alpha \in [0,1]$. $\alpha=0$ for Ito
 convention, while $\alpha=1/2$ and $\alpha=1$ for Stratonovich and and isothermal conventions, 
 respectively. In earlier literature \cite{san82}, the underdamped Langevin equation in a Stratonovich prescription is derived. 
In \cite{lub07}, it has been shown that for all values of $\alpha$, 
the same equilibrium distribution is obtained for a given value of the protocol. Now we closely follow the treatment given in \cite{lub07}.  From 
eq.(\ref{eq:over}), the path probability for a single trajectory in position space can be shown to be given by
\begin{equation}
P[X(t)|x_0]\sim e^{-S[X]},
\end{equation}
where
\begin{equation}
 S[X]=\int_0^{\tau}dt \left( \frac{1}{2g^2}[\dot x - f(x,t) + \alpha gg']^2 +\alpha f'(x,t)\right).
\end{equation}
Using $f(x,t)=-U'(x,t)\Gamma(x) + (1-\alpha)g(x)g'(x)$, we get
\begin{align}
  S[X]= \int_0^{\tau}dt \left( \frac{1}{2g^2}[\dot x + U'\Gamma  + (2\alpha-1) gg']^2\right.\nn\\
\left. +\alpha [-U''\Gamma - U'\Gamma' +(1-\alpha)(gg''+g'^2) ]\right).
\label{ac}
\end{align}
For reverse path, (see eq. (22) of \cite{bar12},\footnote{However, note that in our case, the equilibrium distribution 
is independent of discretization scheme, unlike in \cite{bar12}}) one has to replace $\dot x\rightarrow -\dot x$, and
 $\alpha\rightarrow 1-\alpha$. Thus the action for reverse path is given by,
\begin{small}
\begin{align}
 \tilde S[\tilde X]=\int_0^{\tau}dt \left( \frac{1}{2g^2}[-\dot x - f(x,t) 
+(1- \alpha) gg']^2 \right.\nn\\
 \left.+(1-\alpha) f'(x,t)\right).
\end{align}
\end{small}
Once again, substituting $f(x,t)=-U'(x,t)\Gamma(x) + \alpha g(x)g'(x)$, we get
\begin{align}
  \tilde S[\tilde X]= \int_0^{\tau}dt \left( \frac{1}{2g^2}[-\dot x + U'\Gamma-(2\alpha-1)gg']^2
\right.\nn\\
 \left. +(1-\alpha) [-U''\Gamma - U'\Gamma' +\alpha(gg''+g'^2) ]\right).
\end{align}
However we restrict our analysis to $\alpha=1/2$, i.e, Stratonovich discretization scheme. For this we have
\begin{align}
S[X] =& \int_0^{\tau}dt \left( \frac{1}{2g^2}[\dot x + U'\Gamma]^2\right.\nn\\
&\left. + \frac{1}{2}\left[-U''\Gamma - U'\Gamma' +\frac{1}{2}(gg''+g'^2) \right]\right).
\end{align}
Similarly,
\begin{align}
  \tilde S[\tilde X]= \int_0^{\tau}dt \left( \frac{1}{2g^2}[-\dot x + U'\Gamma  ]^2
\right.\nn\\
 \left. + \frac{1}{2}[-U''\Gamma - U'\Gamma' +\frac{1}{2}(gg''+g'^2) ]\right).
\end{align}
Thus, the path ratio become simply
\begin{align}
\frac{P[X|x_0]}{\tilde P[\tilde X|x_{\tau}]} &= e^{\tilde S[\tilde X]-S[X]} \nn\\
&= \exp\left[-\int_0^\tau dt~ \dot x U'\right] = e^{\beta Q},
\label{eq:mic_rev_ov}
\end{align}
where $Q\equiv -\int_0^\tau dt \dot x U'(x,t) $.
Thus, under Stratonovich scheme, the Crooks fluctuation theorem for trajectories remains unaffected in the overdamped regime, even in the presence of multiplicative noise. Since the Stratonovich scheme is considered to be the physically correct one for a Brownian particle in a heat bath \cite{van},  we may conclude that all the fluctuation theorems retain their forms as in the underdamped case \footnote{The above reasoning is correct for systems where the noise is not exactly delta-correlated, but has a very short correlation time. Now if we take the limit of correlation time going to zero, we get the Fokker-Planck equation that corresponds to the Stratonovich discretization scheme. As is evident, this is the case with most stochastic systems in nature.}.

\begin{figure}
\vspace{0.5cm}
\centering
\includegraphics[width=\lw]{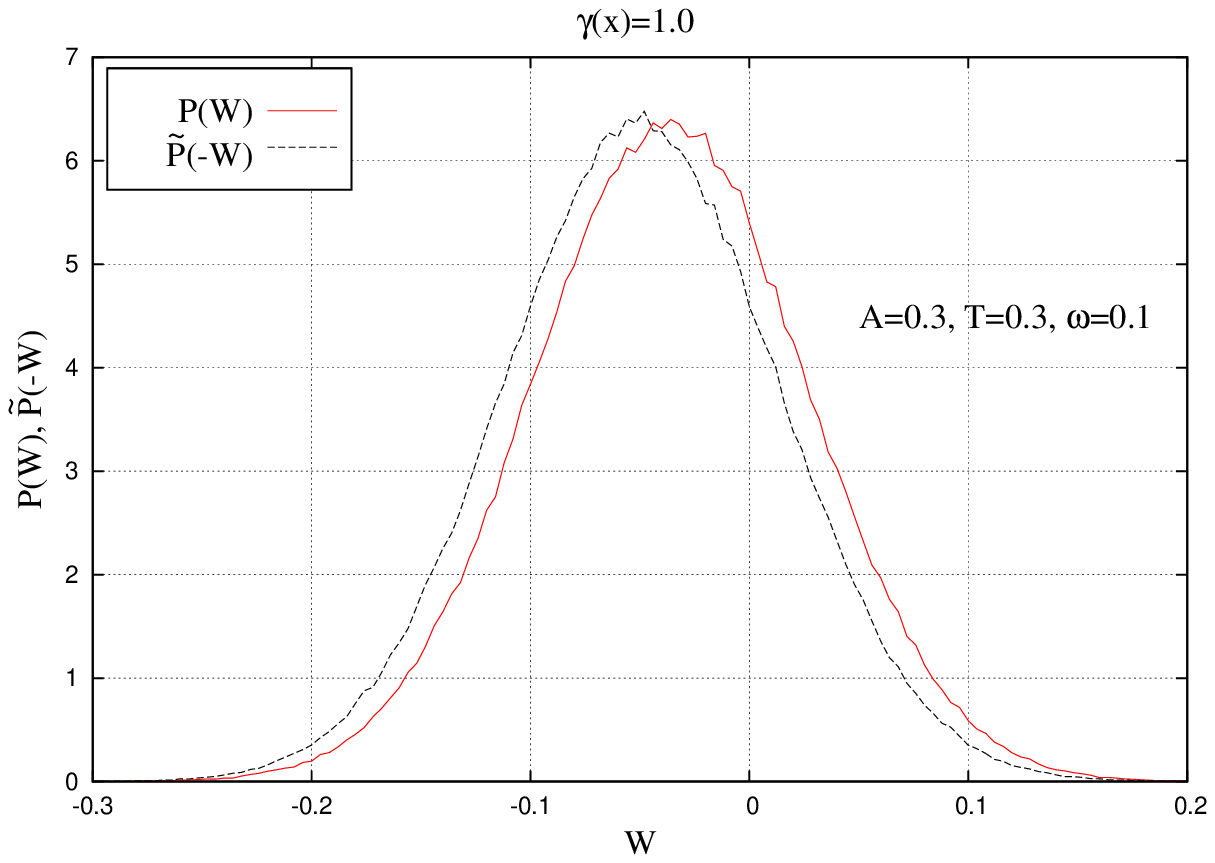}
\caption{Transient work distribution for overdamped case with space-independent friction}
\label{o-w0}
\end{figure}

\begin{figure}
\vspace{0.5cm}
\centering
\includegraphics[width=\lw]{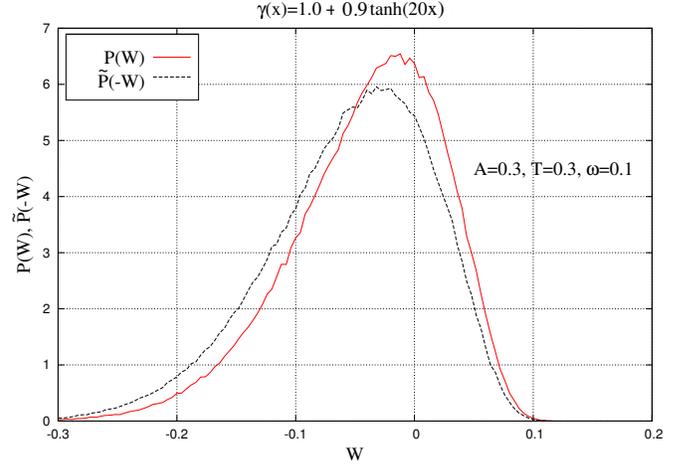}
\caption{Transient work distribution for overdamped case with space dependent friction }
\label{o-wx}
\end{figure}
As in the underdamped case we study the nature of probability distribution for work and entropy for simple model 
of driven harmonic oscillator for both space independent and space dependent case and verifying FTs 
numerically using Heun's method (which is equivalent to following Stratonovich description as discussed
 earlier). The corresponding Langevin equation is given by
\begin{equation}
 \gamma(x)\dot x= -k x +A \sin(\omega t)-\frac{\gamma'(x)}{2\gamma(x)}T+\sqrt{2\gamma (x) T}~\xi(t).
\end{equation}
 In fig.(\ref{o-w0}) and fig.(\ref{o-wx}), we have plotted   the transient work distributions
 for forward and reverse processes, for space-independent and space-dependent
friction, respectively. The functional form of $\gamma(x)$ are same as studied in underdamped case.
All the units are in dimensionless form and we take $k=1$, $\gamma=1$.
 We find that the distributions are Gaussian for space independent
case while for space dependent it is non-Gaussian. But, the crossing point is same. From 
 $\left<  e^{-\beta W}\right> = e^{-\beta\Delta F}$, the numerically obtained free energy difference
$\Delta F=-0.045$, which is  equal to the theoretical value, thus reassuring that space dependent friction 
does not alter the equilibrium distribution.
\begin{figure}
\vspace{0.5cm}
\centering
\includegraphics[width=\lw]{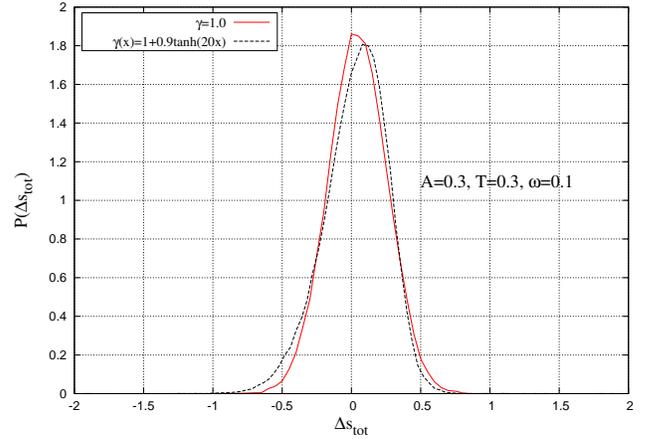}
\caption{Distribution for total entropy production for overdamped dynamics}
\label{ov-s}
\end{figure}

In fig.(\ref{ov-s}) we have plotted the distribution of total entropy production for the overdamped particle. We found that for space independent case the distribution is Gaussian.
This is true only if initial distribution is the thermal one. We have verified separately that for initial nonequilibrium
distribution, $P(\Delta s_{tot})$ is non Gaussian. But for space 
dependent case  $P(\Delta s_{tot})$ is non-Gaussian even initial equilibrium distribution.
 Numerically we find $\la e^{-\Delta s_{tot}}\ra=1.002$ which is well within our numerical accuracy.

\section{Definition of heat in overdamped case}

We can, following Sekimoto \cite{sek}, derive the expression for dissipated heat using the overdamped Langevin dynamics (substituting $\alpha=1/2$ in eq. \eqref{eq:over}):
\begin{align}
\dot x =-\Gamma(x)U'(x,t)+\frac{1}{2}g(x)g'(x) + g(x)\xi(t).
\end{align}
We found that microscopic reversibility gives (see eq. \eqref{eq:mic_rev_ov})
\begin{align}
Q = -\int_0^\tau dt ~\dot x U'(x,t).
\end{align}
The above two equations then give
\begin{align}
Q &= \int_0^\tau dt ~\frac{\dot x}{\Gamma(x)} \left[\dot x - \frac{1}{2}g(x)g'(x) - g(x)\xi(t)\right] \nn\\
&=  \int_0^\tau dt ~\dot x \left[\gamma(x)\dot x + \frac{\gamma'(x)T}{2\gamma(x)}-\sqrt{2\gamma(x)T}~\xi(t)\right] \nn\\
&= \int_0^\tau dt ~\dot x [\gamma(x)\dot x -\sqrt{2\gamma(x)T}~\xi(t)] + \frac{T}{2}\ln \frac{\gamma(x_\tau)}{\gamma(x_0)} \nn\\
&= Q_{conv} + \frac{T}{2}\ln \frac{\gamma(x_\tau)}{\gamma(x_0)},
\end{align}
where $Q_{conv}$ is the conventional definition of heat. We thus get an extra boundary term in the definition, which assigns the logarithm of$\sqrt{\gamma(x)}$ with the physical meaning of an entropy term. The presence of this term implies that if the particle begins from a given position $x_0$ with a small friction coefficient, then it dissipates more heat into the bath if it travels to a position $x_\tau$ with a greater friction coefficient.

\section{Conclusion}
In this work, we have considered the validity of FTs in presence of space-dependent friction, for both
 underdamped and overdamped limit. We find that, although no conceptual difficulties arise when the system is 
underdamped, the derivation of the FTs are more involved for overdamped system. In latter case, where we have dealt with the Stratonovich scheme of discretization, 
the Langevin equation contains extra terms and although Crooks theorem remains valid, the definition of heat gets altered. 
 Thus, we conclude that the FTs remain valid for the case of dynamics of a particle in space 
dependent frictional medium, even under coarse graining, i.e, reducing the description of the system 
of two phase space variable 
(x,v,underdamped case), to a single phase space variable (x,overdamped case). 

As an illustration, we have analyzed the nature of $P(\Delta s_{tot})$ and $P(W)$ for the simple case of a
driven harmonic oscillator in presence of space dependent friction, both in underdamped and the overdamped
regime (in Stratonovich prescription). Distributions for constant friction are compared with that of a 
particle moving in a space dependent frictional medium. Moreover, several FTs have been verified.
This model system is amenable to experimental verification \cite{cil07}.

\section{Acknowledgments}
A.M.J. thanks DST India for financial help. One of us (SL) thanks Prof. Fabio Marchesoni for clarifications
related to the Ito-Stratonovich dilemma and for suggesting useful references.

\end{document}